\documentclass[pra,twocolumn,amsmath,amssymb,showpacs,floatfix]{revtex4}

\usepackage{graphicx}
\usepackage{color}

\newcommand{\refe}[1]{\unskip~(\ref{#1})}
\newcommand{\av}[1]{\langle #1 \rangle}

\renewcommand{\vec}[1]{\mathbf{#1}}
\newlength\figurewidth
\begin{document}

\title{Detecting quadrupole interactions in ultracold Fermi gases}
\begin{abstract}
We propose to detect quadrupole interactions of neutral ultra-cold atoms via their induced mean-field shift.
We consider a Mott insulator state of spin-polarized atoms in a two-dimensional optical square lattice. 
The quadrupole moments of the atoms are aligned by an external magnetic field. As the alignment angle is varied, the mean-field shift shows a characteristic angular dependence, which constitutes the defining signature of the quadrupole interaction. 
  For the $^{3}P_{2}$ states of Yb and Sr atoms, we find a frequency shift of the order of tens of Hertz, which can be realistically detected in experiment with current technology. We compare our results to the mean-field shift of a spin-polarized quasi-2D Fermi gas in continuum.
\end{abstract}  
\author{M.~Lahrz$^{1,2}$, M.~Lemeshko$^{3,4}$, K.~Sengstock$^{1,2,5}$, C.~Becker$^{1,2,5}$, L.~Mathey$^{1,2,5}$}
\affiliation{
\mbox{$^{1}$Zentrum f\"ur Optische Quantentechnologien, 
Universit\"at Hamburg, 22761 Hamburg, Germany}\\
\mbox{$^{2}$Institut f\"ur Laserphysik, Universit\"at Hamburg, 22761 Hamburg, Germany}\\
\mbox{$^{3}$ITAMP, Harvard-Smithsonian Center for Astrophysics, 60 Garden Street, Cambridge, MA 02138, USA}\\
\mbox{$^{4}$Physics Department, Harvard University, 17 Oxford Street, Cambridge, MA 02138, USA}
\mbox{$^{5}$The Hamburg Centre for Ultrafast Imaging, Luruper Chaussee 149, Hamburg 22761, Germany}
}
\date{\today}

\maketitle

%%% INTRODUCTION %%%

\section{Introduction}
The field of ultracold atomic gases has progressed rapidly over the last decades.
This progress was driven by the ability to continuously discover and control new features of these systems~\cite{Lewenstein2007,Bloch2008}.
Optical lattices were used to create reduced dimensions, as well as lattice systems in the strongly interacting limit~\cite{Jaksch1998,Greiner2002}; mixtures of several internal states of ultracold atoms were used to create effective pseudo-spin systems~\cite{Myatt1997,Stenger1998}; the unitary regime of ultracold fermions was explored via the control of bound molecular states via Feshbach resonances~\cite{Greiner2003,Jochim2003,Zwierlein2003}, to name just a few.

%Confining atoms in optical lattices opened up a perspective to study the models of condensed matter physics with different dimensionality; controlling the internal atomic states allowed to provide the particles with an effective pseudo-spin; finally, using Feshbach resonances to tune interatomic interactions paved the way to reaching the strongly-interacting or "unitary" regime. \comment{citations!}

%Optical lattices were used to create reduced dimensions as well as lattice systems in the strongly interacting limit,
%%~\cite{Ospelkaus2008}, 
%mixtures of several internal states of ultracold atoms created effective pseudo-spin systems, and the unitary regime of these systems was explored via    the control of bound molecular states via Feshbach resonances,
%  %~\cite{Land2008,Deiglmayr2008}
%to name just a few. 
In the recent past, atoms and molecules with large electric and magnetic dipole moments were cooled into or near quantum degeneracy~\cite{Koch2008,Ni2008, Aikawa2012,Lu2012}.
 In Ref. \cite{Lahaye2007} the long-range and anisotropic character of the dipole-dipole interaction was demonstrated, 
%The long-range and anisotropic character of the dipole-dipole interaction
 which goes beyond the properties of contact potentials, as discussed in Refs.~\cite{Lahaye2009,Wall2013,Lemeshko2013}.   Theoretical predictions have been made in Refs.~\cite{Baranov2008,Baranov2012,Chan2010,Kestner2010}, that include the formation of supersolids~\cite{Capogrosso2010}, quantum liquid crystals~\cite{Quintanilla2009,Fregoso2009}, and bond-order solids~\cite{Bhongale2012,Bhongale2013b} in dipolar gases. Furthermore, many-body phases in 1D geometries have been reported in Refs.~\cite{Dalmonte2011,Wunsch2011}.

\begin{figure}
\begin{center}
\includegraphics{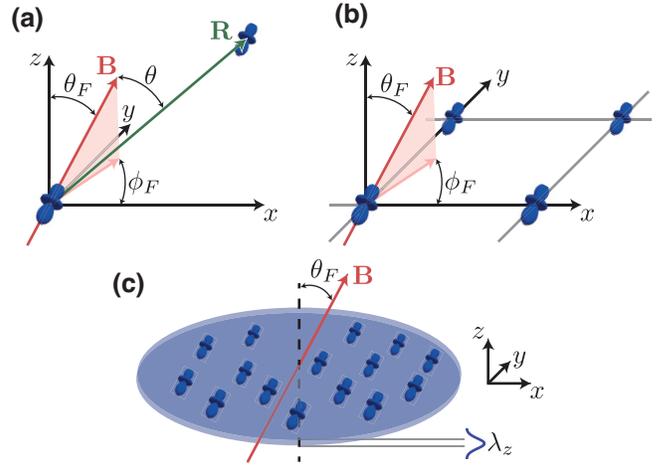}
\caption{(Color online) We consider spin-polarized fermions interacting only via quadrupole-quadrupole interactions. The quadrupoles are aligned by an external field ${\mathbf B}$, that points in the direction defined by the angles $\theta_{F}$ and $\phi_{F}$ as depicted in (a). We propose to measure the mean-field shift of a Mott state in a 2D optical lattice, shown in (b), as a function of the angles $\theta_{F}$ and $\phi_{F}$. For comparison, we also consider a quasi-2D system of width $\lambda_{z}$ of fermions in continuum, panel (c), for which  we determine the mean field shift as a function of  $\theta_F$.}
\label{fig:geometry}
\end{center}
\end{figure}

Here we investigate quadrupole-quadrupole interactions
%Quadrupole-quadrupole interactions have also been established as a peculiar feature of
as a novel feature of ultracold atom systems. 
As pointed out in Ref. \cite{Bhongale2013}, for alkaline-earth atoms such as Yb and Sr in any $M_{J}$ state of the ${}^3P_{2}$ manifold, prepared in an optical lattice with a lattice constant of a few hundreds of nanometers, these interactions are, while small, of realistic magnitude to be relevant in current experiments. 
The quantum phases of quadrupolar quantum gases
 %of these atoms, or of molecules with this type of interaction,
%of alkaline-earth atoms and homonuclear molecules
have been discussed in Refs.~\cite{Bhongale2013, Huang2013}. 
%However, before observing the type of order that the interaction induces, the first step is to demonstrate its existence.
Here, we provide a detailed theoretical study exploring the feasibility of detecting quadrupole-quadrupole interaction in a realistic experiment.

In particular, we propose to detect the quadrupole-quadrupole interaction between ultracold atoms via ultra-high precision
% the mean-field shift that induces.
spectroscopy of the induced mean-field shift.
We consider a Mott insulator state in a 2D optical square lattice
%within one layer of a perpendicular 1D lattice.
and under strong confinement in the third direction.
%As an alternative experimental realization,
 For comparison,
 we study a quasi-2D Fermi gas of spin-polarized atoms in continuum. 
%where we dropped the square lattice
%without a periodic lattice potential but still a confinement in the third direction. 
Both systems are depicted in Fig.~\ref{fig:geometry}, as well as their key parameters.
We assume the  quadrupole axes of all atoms to be aligned by a constant magnetic field $\vec{B}$. 
% by a constant magnetic field that points  in the direction described by the polar angle $\theta_{F}$ and the azimuthal angle $\phi_{F}$, as depicted.
 The atoms are prepared in a state with a particular projection, $M_{J}$, of the angular momentum, $J$, along the quantization axis. The quantization axis is given by the field $\vec{B}$, which points in the direction defined by the polar angles $\left(\theta_F, \phi_F\right)$, shown in Fig.~\ref{fig:geometry}(a).
In the case of the 2D system in continuum, Fig.~\ref{fig:geometry}(c), the angle $\phi_{F}$ is irrelevant, due to the cylindrical symmetry.
%of the system.
As the main example we consider Yb and Sr in a single $M_{J}$ state of the ${}^3P_{2}$ manifold,  where $M_J$ gives the projection of the total angular momentum $J=2$ onto the quantization axis.
%These states are defined with relation to a quantization axis that is given by a magnetic field ${\mathbf B}$.

The angular dependence of the quadrupole-quadrupole interaction potential is the characteristic feature of this interaction. We therefore propose to measure the mean-field frequency shift per particle as a function of $\theta_{F}$ and
%, in the case of the lattice system,
$\phi_{F}$.
Below, we calculate this mean-field shift for both systems, and we
%indeed find a
find a characteristic angular dependence, as shown in Figs. \ref{fig:deltaall}(a) and \ref{fig:continuum}. 
We observe that the interaction changes from repulsive to attractive and back within the range of $0\leq \theta_{F} \leq \pi/2$.
%in the example shown in Fig.~\ref{fig:continuum}.
This characteristic angular dependence constitutes the 'smoking gun' feature to be found experimentally. 
We note that the magnitude of the shift of tens of Hertz for Yb(${}^3P_{2}$) and similarly for Sr(${}^3P_{2}$) makes this experimentally conceivable with current technology.
   
This energy shift has to be measured with respect to a reference state. We propose to use the ground state, ${}^{1}S_{0}$, which has no quadrupole moment and for which the optical transition can be probed~\cite{Yamaguchi2008}. Since we want to verify the quadrupole-quadrupole interaction,
only present in the ${}^{3}P_{2}$ manifold, the initial state for this spectroscopic measurement needs to be the ${}^{3}P_{2}$ state. For this, we propose to first transfer the quantum degenerated atoms to the meta-stable ${}^{3}P_{2}$ state with a short clock-laser pulse. Subsequently a high-resolution spectroscopy is performed by coherently deexciting atoms to the ground state for various angles $\theta_{F}$ of the magnetic field. The quadrupole-quadrupole interaction can be demonstrated with  this characteristic angular dependence.  

A second approach we propose is  radio-frequency (RF) spectroscopy between different $M_{J}$ states of the ${}^{3}P_{2}$ manifold, split by the Zeeman effect, which exhibits different quadrupole moments proportional to $q \propto \left(J^{2}+J-3M_{J}^{2}\right)^{2}$, see Ref.~\cite{Bhongale2013}. For $J = 2$, this gives a ratio between the quadrupole moment $q$ of the $M_{J} \in \left\{-2,0,2\right\}$ states, and of the $M_{J} = \left\{-1,1\right\}$ states, $q^{\prime}$, of $q/q^{\prime} = 4$. Thus, the resulting mean-field energy shift is slightly reduced, however this could be overcompensated by the high resolution of RF spectroscopy.

%In an alternative experiment, the reference state can also be another $M_{J}$ state of the ${}^3P_{2}$ manifold, say $M_{J} = 2$ and $M_{J} = 1$, split by the Zeeman effect. For this, the energy shift of the transition can be measured via radio-frequency (RF) spectroscopy.
%As linear responds the system can be reduced to the populated and its neighboring energy level.
%Here, the reference state has a quadrupole moment, too, since it is proportional to . Thus, the final energy shift will be reduced. However, the high resolution of RF spectroscopy might surpass this counterproductive effect again.

In the Sec. \ref{setup} we introduce the quadrupole-quadrupole interaction and the effective interaction  a quasi-2D geometry. The characteristics of the mean-field frequency shift of a Mott insulator state in a 2D optical square lattice  are discussed in Sec. \ref{lattice}, followed by the treatment of a spin-polarized Fermi gas in continuum in Sec. \ref{gas}. Finally, we conclude in Sec. \ref{conclusion}.

%%% SETUP %%%

\section{General setup}\label{setup}
%
%We consider a system of atoms or molecules, which interact via an electric quadrupole-quadrupole interaction.
We consider a system of atoms possessing an electric quadrupole moment $q$ aligned by an external magnetic field $\vec{B}$.
% The quadrupoles are aligned by an external magnetic field $\vec{B}$, and have an electric quadrupole moment $q$.
 The quadrupole-quadrupole interaction energy between two atoms at positions $\vec{R}_1$ and $\vec{R}_2$, respectively, is
\begin{align}\label{QQI}
U{\left(\vec{R}\right)}
&= \frac{1}{4\pi\varepsilon_{0}}\cdot\frac{3q^2}{16} \cdot\frac{3-30\cos^2{\theta} + 35\cos^4{\theta}}{R^5}
\end{align}
where $\vec{R} \equiv\vec{R}_1-\vec{R}_2$ is the displacement vector, $R\equiv|\vec{R}|$,  and $\theta$ the angle between $\vec{B}$ and $\vec{R}$, cf. Fig.~\ref{fig:geometry}(a). We also introduce the notation $U{\left(R,\theta{\left(\vec{R}\right)}\right)} \equiv U{\left(\vec{R}\right)}$. The interaction given in Eq. \refe{QQI} can also be written in the form $U(R, \theta) = \frac{3}{2} q^2  P_4{\left(\cos\theta\right)}/\left(4\pi\varepsilon_{0} R^{5}\right)$ where $P_4{\left(x\right)}=\left(3-30x^2+35x^4\right)/8$ is the fourth Legendre polynomial.

The magnitudes of the  quadrupole moment $q$ have been reported to be $q \sim 30\,a_{B}^{2}e$ for Yb(${}^3P_{2}$) in Ref.~\cite{Buchachenko-2011} and $q\sim 16\,a_{B}^{2}e$ for Sr(${}^3P_{2}$) in Refs.~\cite{Santra-2004,Derevianko2001}, where $a_{B}$ is the Bohr radius and $e$ the elementary charge. We define the prefactor of Eq. \refe{QQI} as $C_q/\hbar \equiv 3q^2/\left(64\pi\varepsilon_{0}\right)$, which has the dimension of frequency $\times$ length${}^5$. For the states mentioned before, this gives $C_{q}/\hbar \sim 2\pi\times4.59\cdot10^{11}\,\mathrm{Hz\,nm^5}$ and $C_{q}/\hbar \sim 2\pi\times1.31\cdot10^{11}\,\mathrm{Hz\,nm^5}$, respectively.
 Based on Eq. \refe{QQI} we observe that the quadrupole-quadrupole interaction is repulsive for relative angles both $\theta \approx 0$ and $\theta \approx \pi/2$, and attractive at intermediate values. 

\begin{figure}
\begin{center}
\includegraphics[width=\figurewidth]{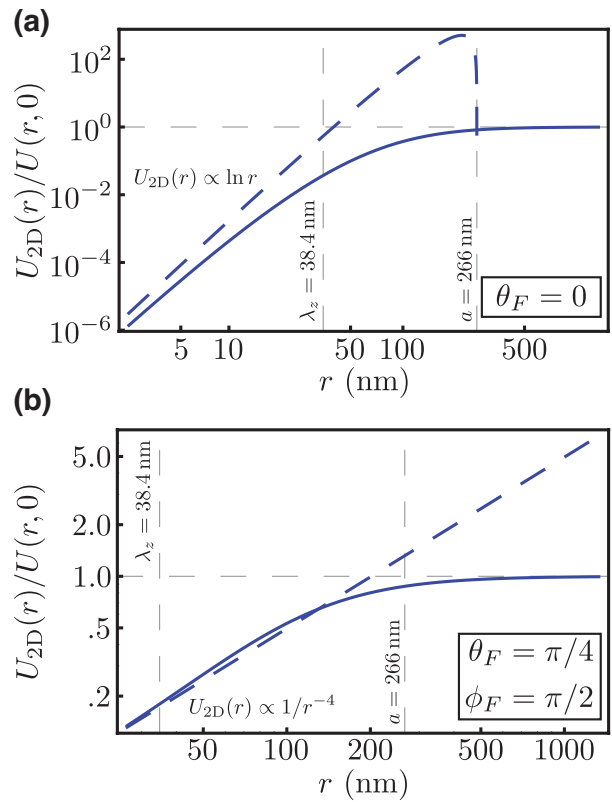}
\caption{(Color online) Effective 2D interaction $U_{\mathrm{2D}}(\vec{r})$, compared to the bare interaction $U(\vec{r})$, for $\lambda_{z} = 38.4$ nm. The spatial extent of the wave function in the $z$-direction provides a cut-off for the effective interaction. 
The ratio of the interactions is depicted by the solid lines, the corresponding short-range limit by the dashed lines.
 (a) The alignment is parallel to the $z$-axis, i.e. $\theta_{F} = 0$ and $\phi_{F}$ is arbitrary. Here the effective interaction is rescaled to a logarithmic behavior. (b) For $\theta_{F} = \pi/4$ and $\phi_{F} = \pi/4$, the resulting effective interaction behaves as $\sim1/r^{4}$ at short distances.}
\label{fig:U2Dcutoff}
\end{center}
\end{figure}

We define the $xy$-plane as the plane of the 2D system.
% In the case of the lattice system, the lattice directions
If an optical lattice is present, its lattice vectors are aligned with the $x$- and $y$-directions of the coordinate system, see Fig.~\ref{fig:geometry}(b). 
 The alignment direction of the quadrupoles is then given by the
 % two angles: $\theta_F$ is the polar angle of the quadrupole direction and $\phi_F$ is the azimuthal angle.
set of polar angles, ($\theta_F$, $\phi_F$). 
With $\vec{R} \equiv \left(x,y,z\right)$,  the alignment angles and the relative angle $\theta$
% appearing in
of Eq. \refe{QQI} are related by 
\begin{align}
\cos{\theta{\left(\vec{R}\right)}} &= \frac{\left(x\cos{\phi_F} + y\sin{\phi_F}\right)\sin{\theta_F} + z\cos{\theta_{F}}}{R}\,.\label{eq:costheta}
\end{align}
We consider the Hamiltonian $\hat{H} = \hat{H}_{0} + \hat{V}$, where
\begin{align}\label{eq:H0}
\hat{H}_{0}  &= \int{\mathrm{d}^3{R}\left(\frac{\left|\nabla \hat{\Psi}^{\dagger}{\left(\vec{R}\right)}\right|^{2}}{2 m} +  \hat{\Psi}^{\dagger}{\left(\vec{R}\right)}\hat{V}_{\mathrm{trap}}\hat{\Psi}{\left(\vec{R}\right)}\right)}
\end{align}
is the Hamiltonian of a single particle in a trapping potential $V_{\mathrm{trap}}$, where $m$ is the atomic mass and $\hat{\Psi}{\left(\vec{R}\right)}$ the fermionic field operator. The interaction $\hat{V}$ is given by
\begin{align}\label{eq:V3d}
\hat{V} &= \frac{1}{2}\int \mathrm{d}^{3}{R}_{1}\mathrm{d}^{3}{R}_{2} U{\left(\vec{R}\right)}
 \hat{\Psi}^{\dagger}{\left(\vec{R}_{1}\right)}\hat{\Psi}^{\dagger}{\left(\vec{R}_{2}\right)}\hat{\Psi}{\left(\vec{R}_{2}\right)}\hat{\Psi}{\left(\vec{R}_{1}\right)}
\end{align}
with $\vec{R} \equiv \vec{R}_{1} - \vec{R}_{2}$.
  The trapping potential separates into $V_{\mathrm{trap}}{\left(\vec{R}\right)} = V_{\mathrm{ol}}{\left(\vec{r}\right)} + V_{\mathrm{c}}{\left(z\right)}$ where $\vec{r}$ is the location in the $xy$-plane defined as $\vec{r} \equiv\left(x,y\right)$. $V_{\mathrm{ol}}{\left(\vec{r}\right)}$ is an optical lattice potential in the $xy$-plane, if present. In the $z$-direction we assume a harmonic confining potential $V_{\mathrm{c}}{\left(z\right)} = m \omega_{z}^{2}z^{2}/2$ with a frequency $\omega_{z}$ and an oscillator length $\lambda_{z} = \sqrt{\hbar/\left(m\omega_z\right)}$. 
  If this confining potential is created by an optical lattice in the $z$-direction, $V_{c}{\left(z\right)} = V_{z}\sin^{2}{\left(\pi z/a_{z}\right)} \approx V_{z}\left(\pi z/a_{z}\right)^{2}$, with 
   lattice constant $a_{z}$ and recoil energy $E_{\mathrm{rec}}^{z} = \hbar^{2}\pi^{2}/(2ma_{z}^{2})$, the oscillator length can be written as $\lambda_{z} = \frac{a_{z}}{\pi}\sqrt[4]{E_{\mathrm{rec}}^{z}/V_{z}}$.
 We assume strong  confinement, and thus in the $z$-direction only the ground state is occupied. We write $\hat{\Psi}{\left(\vec{R}\right)} = \hat{\psi}{\left(\vec{r}\right)}\chi{\left(z\right)}$ where $\chi{\left(z\right)}$ is
\begin{align}\label{eq:chiz}
\chi{\left(z\right)} &= \frac{1}{\left(\pi\lambda_{z}^{2}\right)^{1/4}} \exp{\left(-\frac{z^2}{2\lambda_{z}^2}\right)}\,.
\end{align}
This leads to an effective 2D potential $U_{\mathrm{2D}}(\vec{r})$ by integration out the $z$-component:
\begin{align}\label{eq:U2D}
U_{\mathrm{2D}}(\vec{r}) &= \iint{\mathrm{d}{z}_{1}\mathrm{d}{z_{2}}U{\left(\vec{R}\right)}\left|\chi{\left(z_{1}\right)}\right|^{2} \left|\chi{\left(z_{2}\right)}\right|^{2}}\nonumber\\
&= \int{\mathrm{d}{z}U{\left(\vec{R}\right)}\int{\mathrm{d}{z}_{1}\left|\chi{\left(z_{1}\right)}\right|^{2} \left|\chi{\left(z-z_{1}\right)}\right|^{2}}}\nonumber\\
&= \frac{1}{\left(2\pi\lambda_{z}^{2}\right)^{1/2}}\int{\mathrm{d}{z}U{\left(\vec{R}\right)}\exp{\left(-\frac{z^2}{2\lambda_{z}^2}\right)}}\,,
\end{align}
where $U{\left(\vec{R}\right)}$ is the quadrupole-quadrupole interaction potential of Eq. \refe{QQI}.  For $r/\lambda_{z} \gg 1$ the 3D interaction defined by Eq. \refe{QQI} is reobtained while $\theta$ is given by Eq. \refe{eq:costheta} with $z = 0$. 
 However, $\lambda_{z}$ acts as a cut-off, which in general reduces the power-law behavior to $\sim \left(r/\lambda_{z}\right)^{-4}$ for $r/\lambda_{z}\ll1$.
 For special values of $\phi_F$ or in continuum, it is even further reduced to $\sim\ln{\left(r/\lambda_{z}\right)}$, as shown in Fig.~\ref{fig:U2Dcutoff}. 
 We show the full analytic result, which is depicted in Fig.~\ref{fig:U2Dcutoff}, in the App. \ref{sec:appendix1}.
After integrating out the $z$-direction, as we did in Eq. \refe{eq:U2D}, the effective interaction term $\hat{V}_{\mathrm{2D}}$ becomes
\begin{align}\label{eq:V2D}
\hat{V}_{\mathrm{2D}} &= \frac{1}{2}\int{\mathrm{d}{r}_{1}\mathrm{d}{r}_{2} U_{\mathrm{2D}}{\left(\vec{r}\right)}\hat{\psi}^{\dagger}{\left(\vec{r}_{1}\right)}\hat{\psi}^{\dagger}{\left(\vec{r}_{2}\right)}\hat{\psi}{\left(\vec{r}_{2}\right)}\hat{\psi}{\left(\vec{r}_{1}\right)}}
\end{align}
with $\vec{r} \equiv \vec{r}_{1} - \vec{r}_{2}$.
We propose to measure the energy shift per particle induced by the quadrupole-quadrupole interactions, which is to first order
\begin{align}\label{eq:DE}
\Delta{E} &= \frac{\av{\hat{V}_{\mathrm{2D}}}}{N} = \frac{1}{2n}\int{\mathrm{d}^2{{r}}U_{\mathrm{2D}}{\left(\vec{r}\right)}g{\left(\vec{r}\right)}}
\end{align}
where $n$ is the average density $n = \langle \hat{\psi}^{\dagger}{\left(\vec{r}\right)}\hat{\psi}{\left(\vec{r}\right)}\rangle$, 
\begin{align}\label{eq:g}
g{\left(\vec{r}\right)} &= \frac{1}{A}\int{\mathrm{d}^2{{r}_2}\langle\hat{\psi}^{\dagger}{\left(\vec{r}+\vec{r}_2\right)}\hat{\psi}^{\dagger}{\left(\vec{r}_2\right)}\hat{\psi}{\left(\vec{r}_2\right)}\hat{\psi}{\left(\vec{r}+\vec{r}_2\right)}\rangle}
\end{align}
is the normal-ordered density-density correlation function, and $A$ the system area. In this paper, we refer to $\Delta{E}$ as the mean-field shift.
 We consider a Mott insulator in a 2D optical square lattice in Sec.~\ref{lattice}, and for comparison a spin-polarized Fermi gas in continuum  in Sec.~\ref{gas}.  
%We consider two experimentally conceivable systems in the following, namely a Mott insulator in an optical 2D square lattice in Sec. \ref{lattice} as well as a spin-polarized Fermi gas in a quasi-2D geometry in Sec. \ref{gas}.

%%% LATTICE %%%

\section{Mott state in a 2D optical lattice}\label{lattice}
We consider a Mott insulator state of fermions, which are all in the same internal state, in a deep 2D square lattice in the $xy$-plane, interacting only via quadrupole-quadrupole interactions, Eq. \refe{QQI}.
We assume that the confinement in the $z$-direction is strong enough, so that it can be approximated by a harmonic potential in $z$-direction.
Within the single-band approximation, the field operator can be written as
\begin{align}\label{eq:psiwannier}
\hat{\psi}{\left(\vec{r}\right)}
&= \sum_{i} w{\left(\vec{r} - \vec{r}_{i}\right)} \hat{c}_{i}
%\equiv \sum_{i} w_{i}{\left(\vec{r}\right)} \hat{c}_{i}
\end{align}
where $w(\vec{r})$ is the Wannier function of the lowest band of the lattice and $\hat{c}_{i}$ the corresponding annihilation operator. $\vec{r}_{i}$ are the locations of the lattice minima, and $i$ is the lattice site index. The correlation function of Eq. \refe{eq:g} for the Mott state is 
\begin{align}
g{\left(\vec{r}\right)}
&= n\sum_{i\neq0}{\int{\mathrm{d}^2{{r}_2}}\Big(\left|w{\left(\vec{r}+\vec{r}_2-\vec{r}_{i}\right)}\right|^2\left|w{\left(\vec{r_{2}}\right)}\right|^2}\nonumber\\
&\qquad-
w^{\ast}{\left(\vec{r}+\vec{r}_2-\vec{r}_{i}\right)}w^{\ast}{\left(\vec{r_{2}}\right)}w{\left(\vec{r_{2}-\vec{r}_{i}}\right)}w{\left(\vec{r}+\vec{r}_2\right)}\Big)\label{Mottcorrfunc}
\end{align}
where we used $\langle\hat{c}_{i}^{\dagger}\hat{c}_{j}^{\dagger}\hat{c}_{k}\hat{c}_{l}\rangle = \left(\delta_{il}\delta_{jk} - \delta_{ik}\delta_{jl}\right)\left(1-\delta_{ij}\right)$.

Before we evaluate the correlation function of Eq. \refe{Mottcorrfunc} and $\Delta E$ for the actual Wannier  functions of an optical lattice, we give two simple approximations, which give the correct order of magnitude and the general behavior of $\Delta E$. After that, we evaluate the mean-field shift quantitatively for the correct Wannier states.

\begin{figure}
\begin{center}
\includegraphics[width=\figurewidth]{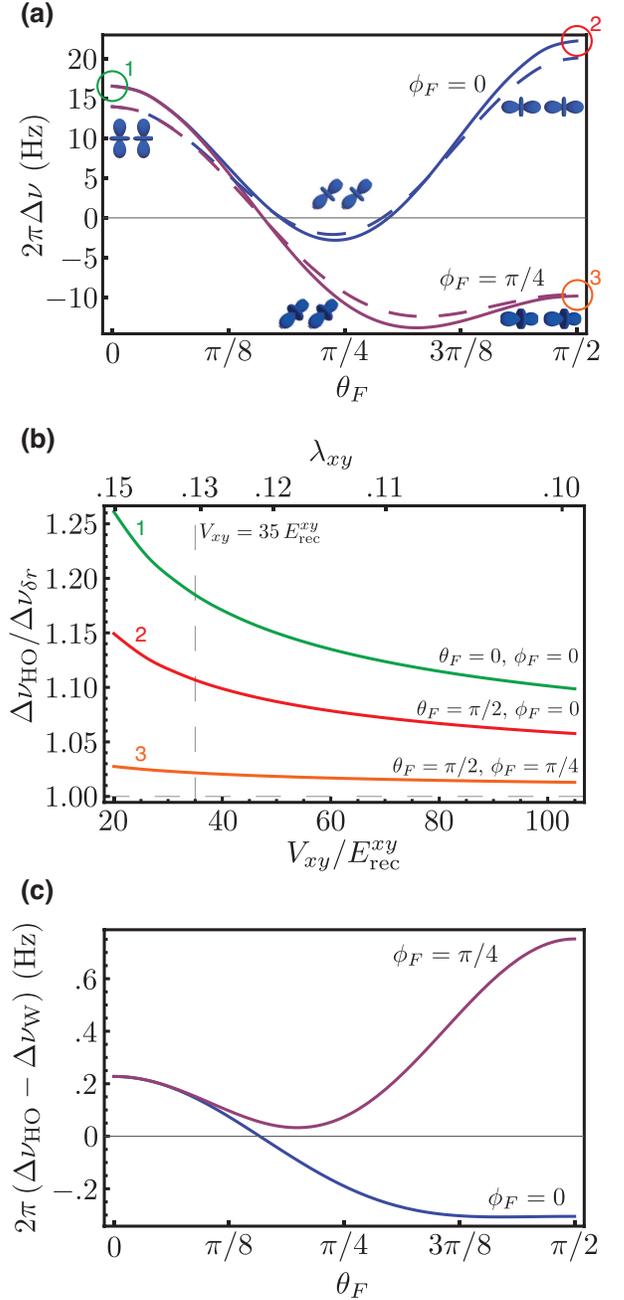}
\caption{(Color online) (a) Mean-field shift for spin-polarized Yb(${}^{3}P_{2}$) atoms in a Mott insulator state, as a function of $\theta_{F}$ and $\phi_{F}$, for the harmonic and point-like approximation of the Wannier states. The solid line corresponds to the case of $\lambda_{z} =\lambda_{xy}= 34.8$ nm, the dashed lines to 
  $\lambda_{z} =34.8$ nm and $\lambda_{xy} \rightarrow 0$. 
 (b) Ratio of the harmonic approximation of the Wannier states to point-like orbitals in the  $xy$-plane, as a function of the lattice depth $V_{xy}/E_{\mathrm{rec}}^{xy}$. The location of $V_{xy} = 35\,E_{\mathrm{rec}}^{xy}$, corresponding to $\lambda_{xy} = 34.8\,\mathrm{nm}$ is marked. (c) Difference in the mean-field frequency shift between the exact Wannier states and the harmonic approximation as a function of $\theta_{F}$ and $\phi_{F}$.}
\label{fig:deltaall}
\end{center}
\end{figure}

We first approximate the Wannier states with harmonic oscillator ground states. Thus, we have a characteristic width $\lambda_{z}$ in $z$-direction, as mentioned above, and a rotationally symmetric oscillator length $\lambda_{xy}$ within the $xy$-plane. This length is much smaller than the lattice constant $a_{xy}$, i.e. $\lambda_{xy} \ll a_{xy}$. The spatial wave function in the $xy$-plane is 
\begin{align}\label{eq:wHO}
w_{\mathrm{HO}}{\left(\vec{r}\right)}
&= \frac{1}{\sqrt{\pi\lambda_{xy}^2}}\exp{\left(-\frac{r^2}{2\lambda_{xy}^2}\right)}\,.
\end{align}
We calculate the correlation function $g{\left(\vec{r}\right)}$ to be 
\begin{align}
g_{\mathrm{HO}}{\left(\vec{r}\right)} &= 
n\sum_{i\neq0}{\frac{1}{2\pi\lambda_{xy}^2}\mathrm{e}^{-\frac{\left(\vec{r}-\vec{r}_{i}\right)^2}{2\lambda_{xy}^2}}\left(1-\mathrm{e}^{-\frac{\vec{r}\cdot\vec{r}_{i}}{\lambda_{xy}^2}}\right)}\,.
\end{align}
As a further approximation, we consider the limit of $\lambda_{xy} \rightarrow 0$.
We continue to assume a finite extent of the Wannier state in $z$-direction, i.e. $\lambda_{z}>0$, but now approximate the Wannier state to be point-like in the $xy$-plane, i.e.  $w_{\delta{r}}{\left(\vec{r}\right)} = \sqrt{\delta{\left(\vec{r}\right)}}$. Then the correlation function reduces to 
\begin{align}\label{eq:gdeltar}
g_{\delta{r}}{\left(\vec{r}\right)} &= n\sum_{i\neq0}{\delta{\left(\vec{r}-\vec{r}_{i}\right)}}\,.
\end{align}
Thus, the energy shift per particle is given by
\begin{align}
\Delta{E}_{\delta{r}} &= \frac{1}{2}\sum_{i\neq0}{U_{2D}{\left(\vec{r}_i\right)}}\,.
\end{align}
In Fig.~\ref{fig:deltaall}(a) we show the mean-field frequency shift $2\pi\Delta{\nu} \equiv \Delta{E}/\hbar$ as a function of $\theta_{F}$ and $\phi_{F}$, for these two approximations,  for Yb(${}^{3}P_{2}$).
 As an example, we consider $\lambda_{xy} = \lambda_{z} = 34.8\,\mathrm{nm}$. This can be achieved with a lattice constant of  $a_{z} = a_{xy} = 266\,\mathrm{nm}$~\cite{Takasu2009,Dorscher2013,Uetake2012} and an optical lattice depth $V_z = V_{xy} = 35\,V_{\mathrm{rec}}$.
 We indeed see a characteristic dependence on the angles $\theta_{F}$ and $\phi_{F}$, and a magnitude of $\sim 10^{1}$ Hz. 
 In Fig.~\ref{fig:deltaall}(b) we show the ratio of these two approximations which is around $\sim 10^{{0}}$. 
 
We now give the full quantitative result, based on the actual Wannier states of a 2D optical square lattice. These are obtained by representing the Hamiltonian \refe{eq:H0} in momentum space, which is then truncated at a high momentum and diagonalized. Out to these eigenstates the Wannier states are constructed by superposition. 
 These Wannier states have the correct overlap of the spatial wave functions of two atoms on different lattice sites which is the main  shortcoming of the harmonic approximation above. Even for the deep lattices considered here, the contribution to the mean-field shift are visible due to the strong interaction at short distances. Note that there is no need for an artificial short-range cut-off in  integral \refe{eq:DE} because the power-law behavior for $r/\lambda_{z}\to0$ is already reduced for the effective 2D potential \refe{eq:U2D}. In addition, the Pauli principle, which is included in Eq. \refe{Mottcorrfunc}, leads to further cancellations.  
  In Fig.~\ref{fig:deltaall}(c) we show the discrepancy between the exact solution, $2\pi\Delta{\nu}_{\mathrm{W}}$, and the harmonic approximation, $2\pi\Delta{\nu}_{\mathrm{HO}}$, with the parameters mentioned above which is of the order of $10^{-1}\,\mathrm{Hz}$. The effect will be even reduced further for stronger confinement in the $x$- and $y$-direction, leading to an increased localization of the atoms on their sites and a reduction of the overlap between neighboring sites.

%%% FERMI GAS %%%

\section{Spin-polarized Fermi gas in a quasi-2D geometry}\label{gas}
In this section, we consider a quasi-2D gas of fermions being all in a single internal state, interacting only via quadrupole-quadrupole interactions. The lattice potential is set to zero, $V_{\mathrm{ol}}{\left(\vec{r}\right)} = 0$. In Ref.~\cite{Yamaguchi2008}, a large inelastic collision rate and  rapid loss of atoms in a dense gas of metastable Yb(${}^{3}P_{2}$) atoms was reported. However, we expect a reduction of the loss rate for the fermionic gas considered here, due to the absence of $s$-wave scattering. 

\begin{figure}[t]
\begin{center}
\includegraphics[width=\figurewidth]{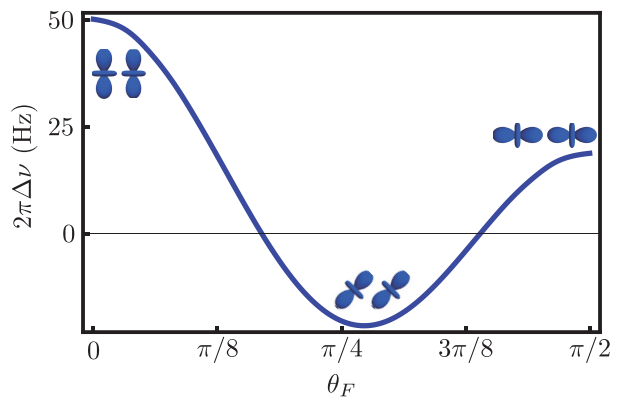}
\caption{(Color online) Characteristic behavior of the mean-field frequency shift $2\pi\Delta{\nu}$ changing  sign twice. 
$\theta_F$ describes the angle of the external relative to the plane. 
 The parameters for this example are given in Sec.~IV. }
\label{fig:continuum}
\end{center}
\end{figure}

The gas is strongly confined in the $z$-direction by a harmonic potential, an the atoms are constrained to the ground state given in Eq. \refe{eq:chiz}. The field operator is
\begin{align}
\hat{\psi}{\left(\vec{r}\right)} &= \frac{1}{\sqrt{A}}\sum_{\vec{k}}{\mathrm{e}^{\mathrm{i}\vec{k}\cdot\vec{r}}\hat{\psi}_{\vec{k}}}\,,
\end{align}
where $A$ is the area of the system. 
 We sum over all wave vectors $\vec{k}$, and $\hat{\psi}_{\vec{k}}$ is the corresponding field operator for the mode $\vec{k}$.
 The correlation function of the non-interacting gas is
\begin{align}\label{eq:gcon}
g_{\mathrm{con}}{\left(r\right)} &= n^{2}\left(1 - \frac{1}{N} \sum_{\vec{k}}{\mathrm{e}^{\mathrm{i}kr\cos{\eta}}n_{k}}\right)\nonumber\\
&= n^{2}\left(1 - \frac{1}{n} \int_{0}^{\infty}{\frac{\mathrm{d}{k}}{2\pi}k\mathrm{J}_{0}{\left( kr\right)}n_{k}}\right)\,,
\end{align}
where $\cos{\eta} = \vec{k}\cdot\vec{r}/\left(kr\right)$.  $n_{k} \equiv \av{\hat{\psi}_{k}^{\dagger}\hat{\psi}_{k}}$ is Fermi distributed according to
\begin{align}
n_{k}
&= \left(Z^{-1}\mathrm{e}^{\frac{E_{k}}{k_{B}T}}+1\right)^{-1}
\end{align}
where $k_{B}$ is the Boltzmann constant, $Z$ is the fugacity, which for a 2D Fermi gas is $Z = \mathrm{e}^{\mu/k_{B}T} = \mathrm{e}^{T_{\mathrm{F}}/T}-1$, and $\mu$ is the chemical potential. $T_{\mathrm{F}} = \hbar^{2}k_{F}^{2}/\left(2mk_{B}\right)$ is the Fermi temperature and $k_{F} = \sqrt{4\pi n}$ is the Fermi vector. 
Note that the pair correlation function only depends on the radial component $r$ and not on the orientation $\alpha = \arg{\left(\vec{r}\right)}$ anymore. Thus, we integrate out this degree of freedom in Eq. \refe{eq:U2D} which gives
\begin{align}\label{eq:U2Da}
U_{\mathrm{2D}}^{\alpha}{\left(r\right)}
&=\frac{\sqrt{2\pi}}{384\lambda_{z}^5}\mathrm{e}^{\frac{r^2}{4\lambda_{z}^2}}\left(20\cos{\left(2\theta_F\right)}+35\cos{\left(4\theta_F\right)}+9\right)
\nonumber\\
&\qquad\times\left[
\left(6 + 6\frac{r^{2}}{\lambda_{z}^{2}} + \frac{r^{4}}{\lambda_{z}^{4}}\right)\mathrm{K}_{0}{\left(\frac{r^{2}}{4\lambda_{z}^{2}}\right)}
\right.
\nonumber\\
&\qquad\quad\left.
-\frac{r^{2}}{\lambda_{z}^{2}}\left(4 + \frac{r^{2}}{\lambda_{z}^{2}}\right)\mathrm{K}_{1}{\left(\frac{r^{2}}{\lambda_{z}^{2}}\right)}
\right]\,.
\end{align}
 $g_{\mathrm{con}}{\left(r\right)}$ is independent of the orientation of the $\vec{B}$-field. Therefore, the angular dependence of the mean-field energy shift
\begin{align}%\label{eq:DE}
\Delta{E} &= \frac{1}{2n}\int{\mathrm{d}{r}rU_{\mathrm{2D}}^{\alpha}{\left(r\right)}g_{\mathrm{con}}{\left(r\right)}}
\end{align}
 can be read off Eq. \refe{eq:U2Da} directly. All other parameters, such as density and quadrupole moment only influence the magnitude of this dependence, cf. Fig.~\ref{fig:continuum}.

For the following numerical examples, we again choose the depth of the trapping potential to be $V_{z} = 35\,V_{\mathrm{rec}}^{z}$  which gives $\lambda_{z} = 0.13\,a_{z}$, and $a_{z} = 266\,\mathrm{nm}$. The quadrupole moment is assumed to be $q \sim 30\,a_{B}^{2}e$ as for Yb(${}^{3}P_{2}$)~\cite{Buchachenko-2011}. For  the 2D density of the gas we choose  $n = 14.1\,\mathrm{\mu m}^{-2}$ which corresponds to a mean 2D interparticle distance of $d =  1/\sqrt{n} =266\,\mathrm{nm}$ which is equal to the lattice constant above.

\begin{figure}[t]
\begin{center}
\includegraphics[width=\figurewidth]{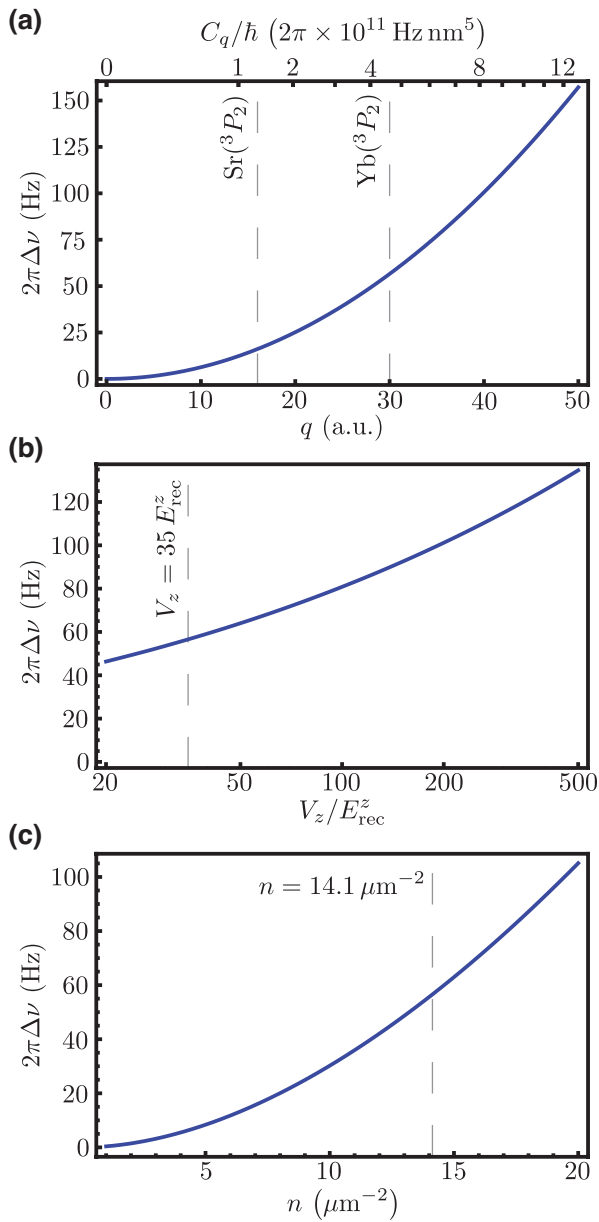}
\caption{(Color online) Frequency shift $2\pi\Delta{\nu}$ for $\theta_{F} = 0$ depending on experimental parameters (a) quadrupole moment, (b) confinement in $z$-direction, and (c) particle density. The parameters for this example are $V_{z} = 35\,E_{\mathrm{rec}}^{z}$, $n = 14.1\,\mathrm{\mu m}^{-2}$, $T = 0$ and $q = 30\,a_{B}^{2}e$ for Yb(${}^{3}P_{2}$), or marked in the corresponding plots, respectively. Additionally, $q = 16\,a_{B}^{2}e$ corresponding to Sr(${}^{3}P_{2}$) is specified in (a).}
\label{fig:parameters}
\end{center}
\end{figure}

In Fig.~\ref{fig:parameters} we show several dependencies of the mean-field shift. In Fig.~\ref{fig:parameters}(a) the dependence on
 on the quadrupole moment $q$ is shown, in  Fig.~\ref{fig:parameters}(b) the dependence on the trapping energy $V_{z}$. 
  In Fig.~\ref{fig:parameters}(c) we show the dependence of the mean-field energy 
 on the particle density $n$.

\begin{figure}
\begin{center}
\includegraphics[width=\figurewidth]{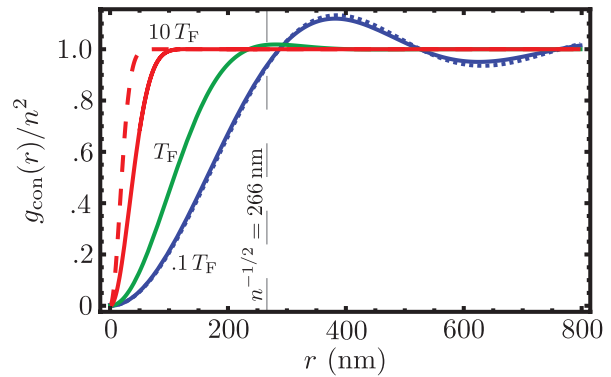}
\caption{(Color online)  Density-density correlation function $g_{\mathrm{con}}{\left(r\right)}$ for different temperatures $T/T_{\mathrm{F}} = 0.1,\,1,\,10$ and $n=14.1\,\mathrm{\mu m}^{-2}$ (solid lines). Analytic results for $T = 0$ (dotted line) and high temperature limit for $T/T_{\mathrm{F}} = 10$ (dashed line).}
\label{fig:g}
\end{center}
\end{figure}
\begin{figure}
\begin{center}
\vspace*{1em}
\includegraphics[width=\figurewidth]{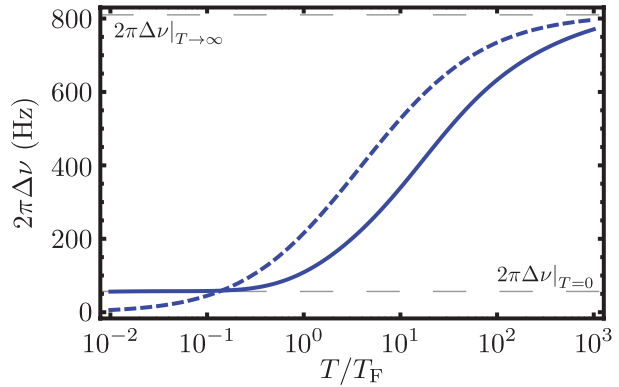}
\caption{(Color online) Mean-field shift of spin-polarized Yb(${}^{3}P_{2}$) atoms as a function of temperature $T/T_{\mathrm{F}}$, for a quasi 2D Fermi gas. The solid line corresponds to the exact result while the dashed line to the high temperature approximation $T/T_{\mathrm{F}}\gg1$. The analytic limiting cases for $T = 0$ and $T/T_{\mathrm{F}} \to \infty$ are marked.
 We use  $V_{z} = 35\,E_{\mathrm{rec}}^{z}$, $n=14.1\,\mathrm{\mu m}$ and $q = 30\,a_{B}^{2}e$ in this example. }
\label{fig:T}
\end{center}
\end{figure}
Finally, the mean-field frequency shift is influenced by the temperature $T$.
The correlation function $g_{\mathrm{con}}{\left(r\right)}$, shown in Fig.~\ref{fig:g}, can be determined analytically for two limiting cases. For zero temperature we find 
\begin{align}
g_{\mathrm{con}}^{T=0}{\left(r\right)}
&= n^{2}\left(1 - 2\frac{\mathrm{J}_{1}{\left(k_{F}r\right)}}{k_{F}r}\right)\,.
\end{align}
For temperatures much larger than the Fermi temperature, i.e. $T / T_{\mathrm{F}} \gg 1$, we find
\begin{align}
g_{\mathrm{con}}^{T/T_{\mathrm{F}}\gg1}{\left(r\right)}
&= n^{2}\left(1 - \mathrm{e}^{-\frac{1}{4}\frac{T}{T_{\mathrm{F}}}k_{F}^{2}r^{2}}\right)\,.
\end{align}
In the limit $T/T_{\mathrm{F}} \to \infty$, the density-density correlations function approaches $g_{\mathrm{con}}^{T/T_{\mathrm{F}}\to\infty}{\left(r\right)} = n^{2}$, for all $r\neq 0$. Thus, the mean-field frequency shift approaches $2\pi\Delta{\nu} = \Delta{E}/\hbar = \frac{n}{2\hbar}\int_{0}^{\infty}{\mathrm{d}{r}rU_{\mathrm{2D}}^{\alpha}{\left(r\right)}}$ which gives an upper limit. In Fig.~\ref{fig:T}, we show the mean-field frequency shift for different temperatures.

%%% CONCLUSION %%%

\section{Conclusion}\label{conclusion}
 In conclusion, we have given a concrete proposal to detect a novel feature of ultra-cold atom systems, quadrupolar interactions. 
 We have proposed to detect these via the mean-field shift they induce, and by its characteristic angular dependence. We have
 demonstrated that the mean-field shift is of realistic magnitude, on the order of tens of Hertz, which can be detected with current technology. For a Mott insulator state in a deep 2D optical lattice  we have found a highly tunable system, in which the mean-field frequency shift can be manipulated with an external magnetic field which controls the  the alignment of the quadrupoles. We also considered a spin-polarized Fermi gas in continuum, for which we found a characteristic dependence on the alignment angle $\theta_{F}$, which is  universal for any experimental realization. Furthermore, we discussed that the scale of the frequency shift is affected by  several experimental parameters such as the quadrupole moment, the confinement in $z$-direction, the particle density and temperature.  We also expect a mean-field frequency shift in a comparable system of molecules to be of the same order of magnitude.

%% acknowledgments %%
\begin{acknowledgments}
We gratefully acknowledge discussions with Florian Schreck, Yoshiro Takahashi, Wen-Min Huang, Eite Tiesinga and Alexander Pikovski.
 We acknowledge support from the Deutsche Forschungsgemeinschaft through
the SFB 925 and the Hamburg Centre for Ultrafast Imaging, and from the Landesexzellenzinitiative Hamburg, which is supported by the Joachim Herz Stiftung.  
\end{acknowledgments}

%%% REFERENCES %%%

%\twocolumngrid

%%% APPENDIX %%%

\onecolumngrid
\appendix
\section{An analytic solution for the effective 2D potential}\label{sec:appendix1}
Here, we show the analytic result for Eq. \refe{eq:U2D}. $U{\left(\vec{R}\right)}$, given by Eq. \refe{QQI}, depends on $z$ via $R = \sqrt{r^{2} + z^{2}}$ and $\cos(\theta{\left(\vec{R}\right)})$ is given by Eq. \refe{eq:costheta}. We find
\begin{small}
\begin{align}
U_{\mathrm{2D}}{\left(\vec{r}\right)}
&= \frac{1}{384\sqrt{2\pi}\lambda_{z}^5}\frac{\lambda_{z}^2}{r^2}\mathrm{e}^{\frac{r^2}{4\lambda_{z}^2}}\nonumber\\
&\quad\times\left\{
\frac{r^2}{\lambda_{z}^2}\mathrm{K}_{0}{\left(\frac{r^2}{4\lambda_{z}^2}\right)}
\left[
-16\left(\frac{r^4}{\lambda_{z}^4}+4\frac{r^2}{\lambda_{z}^2}\right)\sin^2{\left(\theta_F\right)}\left(7\cos{\left(2\theta_F\right)}+5\right)\cos{\left(2\left(\alpha-\phi_F\right)\right)}
\right.\right.
\nonumber\\
&\quad\quad\left.
+8\left(\frac{r^4}{\lambda_{z}^4}-2\frac{r^2}{\lambda_{z}^2}+6\right)\sin^4{\left(\theta_F\right)}\cos{\left(4\left(\alpha-\phi_F\right)\right)}
+\left(\frac{r^4}{\lambda_{z}^4}+6\frac{r^2}{\lambda_{z}^2}+6\right)\left(20\cos{\left(2\theta_F\right)}+35\cos{\left(4\theta_F\right)}+9\right)
\right]
\nonumber\\
&\quad+\mathrm{K}_{1}{\left(\frac{r^2}{4\lambda_{z}^2}\right)}
\left[
16\left(\frac{r^6}{\lambda_{z}^6}+2\frac{r^4}{\lambda_{z}^4}-2\frac{r^2}{\lambda_{z}^2}\right)\sin^2{\left(\theta_F\right)}\left(7\cos{\left(2\theta_F\right)}+5\right)\cos{\left(2\left(\alpha-\phi_F\right)\right)}
\right.
\nonumber\\
&\quad\quad\left.\left.
-8\left(\frac{r^6}{\lambda_{z}^6}-4\frac{r^4}{\lambda_{z}^4}+16\frac{r^2}{\lambda_{z}^2}-48\right)\sin^4{\left(\theta_F\right)}\cos{\left(4\left(\alpha-\phi_F\right)\right)}
+\left(\frac{r^6}{\lambda_{z}^6}+4\frac{r^4}{\lambda_{z}^4}\right)\left(20\cos{\left(2\theta_F\right)}+35\cos{\left(4\theta_F\right)}+9\right)
\right]
\right\}\,,
\end{align}%
\end{small}%
where $r = \left|\vec{r}\right|$, $\alpha = \arg{\left(\vec{r}\right)}$, and $K_{\nu}{\left(x\right)}$ are the modified Bessel functions of the second kind. In both systems that we consider, the square lattice and the continuum, we find a 4-fold rotational symmetry, which implies that all terms proportional to $\cos{\left(2\left(\alpha-\phi_{F}\right)\right)}$ vanish later on in the calculation. For the limit $r/\lambda_{z}\to0$ we find 
\begin{align}
\left.U_{\mathrm{2D}}{\left(\vec{r}\right)}\right|_{\theta_{F}\neq0}
&\overset{r/\lambda_{z}\to0}{=}
\frac{1}{\sqrt{2\pi}\lambda_{z}^{5}}\left(\frac{r}{\lambda_{z}}\right)^{-4} + \mathcal{O}{\left(\frac{r}{\lambda_{z}}\right)^{-3}}\,,
\end{align}
for $\theta_{F} \neq 0$, and
\begin{align}
\left.U_{\mathrm{2D}}{\left(\vec{r}\right)}\right|_{\theta_{F}=0}
&\overset{r/\lambda_{z}\to0}{=}
-\frac{2}{\sqrt{2\pi}\lambda_{z}^{5}} \ln{\left(\frac{r}{\lambda_{z}}\right)} + \mathcal{O}{\left(1\right)}
\end{align}
for $\theta_{F} = 0$ and in continuum, respectively.

\section{Influence of the trapping depths in the point particle limit}\label{sec:appendix2}
%
%\twocolumngrid
%
In Sec. III we introduced the harmonic oscillator lengths $\lambda_{z}$ and $\lambda_{xy}$, see Eqs. \refe{eq:chiz} and \refe{eq:wHO}, respectively. In an experiment these are controlled by the depth of the trapping potential, $V_{\nu}/E_{\mathrm{rec}}^{\nu}$, and the lattice constant, $a_{\nu}$, as $\lambda_{\nu} = a_{\nu}\sqrt[4]{E_{\mathrm{rec}}^{\nu}/V_{\nu}}/\pi$ where $\nu = \left\{z,xy\right\}$. 

\begin{figure}
\begin{center}
\includegraphics[width=\figurewidth]{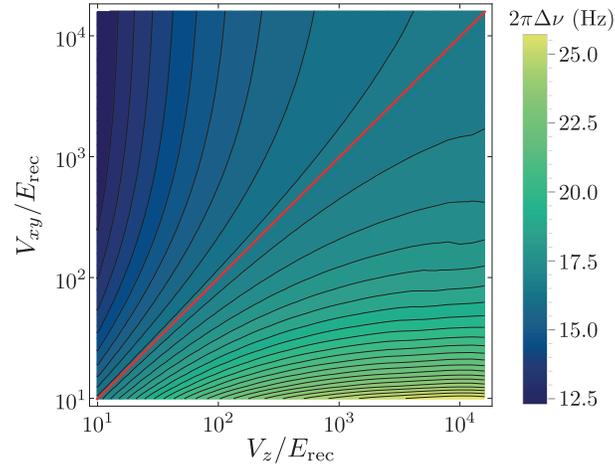}
\caption{(Color online) Mean-field shift of spin-polarized Yb(${}^{3}P_{2}$) atoms in a 2D Mott state as a function of the lattice depths $V_{xy}/E_{rec}$  and $V_{z}/E_{rec}$ in $z-$ and $xy$-direction for equal lattice constants $a_{z} = a_{xy} = 266\,\mathrm{nm}$. The red line corresponds to $V_{z} = V_{xy}$, along which the mean-field shift is almost constant.}
\label{fig:lambda}
\end{center}
\end{figure}

The lattice constant is given by half of the trapping laser wave length which we assume to be the same for all directions $\lambda_{\mathrm{Laser}}/2 = a_{z} = a_{xy} = 266\,\mathrm{nm}$. Note that also $E_{\mathrm{rec}}^{z} = E_{\mathrm{rec}}^{xy}$ since $E_{\mathrm{rec}}^{\nu} = \hbar^{2}\pi^{2}/\left(2ma_{\nu}^{2}\right)$.  In Fig.~\ref{fig:lambda} we show the mean-field shift as a function of $V_{z}/E_{\mathrm{rec}}^{z}$ and $V_{xy}/E_{\mathrm{rec}}^{xy}$.
The mean-field shift becomes smaller if the trapping in $z$-direction is weaker and larger if it is stronger. In fact, for an infinitely deep trap, $V_{z}\to\infty$, the mean-field shift diverges , except  for the limit $V_{xy}\to\infty$ while keeping $V_{z}/V_{xy}\leq1$. This  limit is correctly fulfilled by $\Psi{\left(\vec{R}\right)} = \sqrt{\delta{\left(\vec{R}\right)}}$ and leads to the simple expression $\Delta{E} = \sum_{i\neq0}{U{\left(\vec{R}_{i}\right)}}/2$, where $\vec{R}_{i} = \left(\vec{r}_{i},0\right)$. Interestingly, the mean-field shift is almost constant if $V_{z} = V_{xy}$ making the unsophisticated ansatz of a three-dimensional dot-like spatial wave function surprisingly effective.

\end{document}